# GAW – An Imaging Atmospheric Cherenkov Telescope with Large Field of View

G. Cusumano [1], G. Agnetta [1], A. Alberdi [2], M. Alvarez [3], P. Assis [4], B. Biondo [1], F. Bocchino [5],
P. Brogueira [4],  J.A. Caballero ([3], M. Carvajal [6], A.J. Castro-Tirado [2], O. Catalano [1], F. Celi [1],
C. Delgado [7], G. Di Cocco [8], A. Dominguez [3], J.M. Espino Navas [3], M.C. Espirito Santo [4],
M.I. Gallardo [3], J.E. Garcia [6], S. Giarrusso [1], M. Gomez [6], J.L. Gomez [2], P. Goncalves [4],
M. Guerriero [2], A. La Barbera [1], G. La Rosa [1], M. Lozano [3], M.C. Maccarone [1], A. Mangano [1],
I. Martel [6], E. Massaro [9], T. Mineo [1], M. Moles [2], F. Perez-Bernal [6], M.A. Peres-Torres [2],
M. Pimenta [4], A. Pina [4], F. Prada [2], J.M.  Quesada [3], J.M. Quintana [2], J.R. Quintero [6],
J. Rodriguez [6], F. Russo [1], B. Sacco [1], M.A. Sanchez-Conde [2], A. Segreto [1], B. Tome' [4],
A. de Ugarte Postigo [2], P. Vallania [10]

[1]     IASF-Palermo/INAF, Istituto di Astrofisica Spaziale e Fisica Cosmica di Palermo, Italy
[2]     IAA-CSIC, Istituto de Astrofisica de Andalucia, Granada, Spain
[3]     DFAMN Department, Universida de Sevilla, Spain
[4]     Lab. Instrumentacao e Fìsica de Partìculas, LIP, Av. Elias Garcia 14, 1000-149, Lisbon, Portugal
[5]     OA-Palermo/INAF, Osservatorio Astronomico di Palermo, Italy
[6]     Universidad de Huelva, Spain
[7]     IAC, Istituto de Astrofisica de Canarias, Tenerife, Spain
[8]     IASF-Bologna/INAF, Istituto di Astrofisica Spaziale e Fisica Cosmica di Bologna, Italy
[9]     Physics Dept., Università La Sapienza, Roma, Italy
[10]   IFSI-Torino/INAF, Istituto di Fisica dello Spazio Interplanetario di Torino, Italy

**Abstract**
GAW, acronym for Gamma Air Watch, is a Research and Development experiment in the TeV range, whose main goal is to explore the feasibility of large field of view Imaging Atmospheric Cherenkov Telescopes. GAW is an array of three relatively small telescopes (2.13 m diameter) which differs from the existing and presently planned projects in  two main features: the adoption of a refractive optics system as light collector and the use of single photoelectron counting as detector working mode. The optics system allows to achieve a large field of view (24°x24°) suitable for surveys of large sky regions.  The single photoelectron counting mode in comparison with the charge integration mode improves the sensitivity by permitting also the reconstruction of events with a small number of collected Cherenkov photons. GAW, which is a collaboration effort of Research Institutes in Italy, Portugal and Spain, will be erected in the Calar  Alto Observatory (Sierra de Los Filabres - Andalucia, Spain), at 2150  m a.s.l.). The first telescope will be settled within Autumn 2007.  This paper shows  the main characteristics of the experiment and its expected performance.

**Introduction**
In the last fifteen years a new electromagnetic window (50 GeV - 20 TeV) has been opened thanks to the observation of the sky by ground-based Atmospheric Cherenkov Telescopes (ACT). The detection of the Cherenkov light produced by the electrophotonic showers, originated by the interaction of gamma-ray photons in the atmosphere, has offered spectacular breakthroughs in this extreme observational energy domain. A remarkable number of sources has been firmly detected since the pioneer detection of Very High Energy (VHE) emission from the Crab Nebula [1]. Among the most exciting recent results we cite the discovery of many new sources in the Galactic plane, only some of them identified with known astronomical sources, and the detection of some Active Galactic Nuclei (AGN), three of them with redshifts in the range from 0.15 to 0.2. These are the three most distant extragalactic sources observed in this energy band and their detection is relevant for the evaluation of the Extragalactic Background Light and shows that the intergalactic space is more transparent to gamma-rays than previously thought.
An increasing number of VHE experiments are now exploring the sky to capture emissions in this extreme energy band. They are producing observational results useful for the understanding of the physical processes responsible for the emission in AGNs and supernova remnants and helped us to study the cosmic-ray acceleration processes. The majority of telescopes have imaging capability (IACT), with the Cherenkov radiation focused onto a pixellated camera and the dominant cosmic-ray background is rejected by exploiting differences in the images from primaries  and proton air showers [2]. The main advantages of the atmospheric Cherenkov technique are: high sensitivity, good angular resolution, moderate energy resolution and a low energy threshold that is getting closer and closer to the observational energy window of the gamma-ray space experiments (EGRET, AGILE, GLAST-LAT). IACT, however, have a reduced duty cycle of about 10% and a small Field of View (FoV) of a few degree (3-5 degrees). The duty cycle of ACTs depends upon the constrain of moonless and clean nights, while their limited FoV is tied to the optical telescope design adopted to collect and image the light. Present IACTs , using large mirror reflectors, cannot reach larger FoV because of the mirror optical aberrations, rapidly increasing with off-axis angles. Moreover, the increasing of the detector area necessary to cover large FoV would inevitably produce a strong reduction of the light collecting area of the primary mirror because of the shadow of the focal plane instrumentation onto the reflecting surface.

A large FoV, however, is an important instrumental requirement for several scientific goals. VHE astronomical events, in fact, can occur at unknown locations and/or random in time and a large FoV is then mandatory to increase their detection probability. Moreover, it is also very useful to perform sensitive surveys in and out of the Galactic Plane and to measure the celestial gamma-ray diffuse emission as well. A further advantage offered by a large FoV is the increase of the effective area at the highest energies: the more distant is the core of the Cherenkov light pool from the telescope, the farther the image falls from the center of the focal plane detector. Large FoV will keep these events inside the detector area, increasing the sensitivity of the telescope for the most energetic electromagnetic radiation.

In this paper we describe the main characteristics of GAW, acronym for Gamma Air Watch, a R&D experiment that will test the feasibility of a new generation of IACT that incorporates high flux sensitivity with large FoV capability, stereoscopic observational approach and single photoelectron counting mode. A technical description of the GAW experiment can be found in other two papers presented at this conference [3, 4].

## The experiment

GAW is conceived as an array composed by three identical IACTs located at the vertexes of a triangle, ~80 m side. A detailed description of GAW is given in [5]

GAW is different from the present IACT experiments.

A refractive optical system characterizes its optics system: the light collector is a non commercial single side flat Fresnel lens (2.13 m diameter) with focal length of 2.55 m and thickness of 3.2 mm. The Fresnel lens, an approximation of the refractive aspherical lens, provides a large FoV with imaging quality suitable to the coarse structure of the Cherenkov image and with the advantage of no central obstruction of the focal plane detector. The lens is made of UltraViolet (UV) transmitting acrylic with a nominal transmittance of ~95% from 330 nm to the near InfraRed. The material has a small refraction index derivative at low wavelength, thus reducing chromatic aberration effect. This will be further minimized by implementing diffractive optics design onto the side of the Fresnel lens containing the grooves. The lens design is optimized at ~360 nm, and it is characterized by a quite uniform spatial resolution suitable to the requirements of the Cherenkov imaging up to 12° off-axis. The lens is made of 33 petals maintained in a rigid configuration by a spider structure. The optical system is designed and manufactured by the Fresnel Technologies, Fort Worth, Texas.

Another important difference of GAW with respect to traditional Cherenkov telescopes is the detector working mode. The focal plane detector of each telescope consists of a grid of 40x40 Multi-Anode Photomultipliers Tubes (MAPMT), with 64 anodes each, arranged in 8x8 matrix, operated in single photoelectron counting mode [6] instead of the charge integration method widely used in the IACT experiments. The active channels, 102400 for each telescope, will record the Cherenkov binary image with high granularity, which is fundamental to minimize the probability of photoelectron pile-up within intervals shorter than the sampling time of 10 ns. In such working mode, the electronic noise and the PMT gain differences are negligible, allowing to lower the photoelectron trigger threshold and, as a consequent result, to achieve a low telescope energy threshold in spite of the relatively small dimension of the Cherenkov light-collector. The stereoscopic observational approach will guarantee the sensitivity necessary to accomplish its feasibility and scientific goals.

GAW will be erected in the Calar Alto Observatory (Sierra de Los Filabres - Andalucía, Spain), at the altitude of 2150 m a.s.l.. Two phases are foreseen for the project:

in the first phase, only part of the GAW focal detector will be implemented to cover a FoV of about 5°x5°. The detector will be mounted on a rack frame and it can be moved to cover the entire focal plane. The instrumental sensitivity in this configuration will be tested observing the Crab Nebula on- and off-axis up to 12°.

In the second phase the focal plane detector will be enlarged to cover a FoV of 24°x24°. We plan to survey a sky belt with an extension of  in the North-South direction.

## GAW expected performances

The GAW expected performances were evaluated with a complete end-to-end simulation. Physical processes involved in the interaction of a gamma-ray or a proton in the atmosphere, shower production and development, generation of Cherenkov light and effects of the atmospheric absorption were simulated using the CORSIKA code [7]. The collection of the Cherenkov light by the stereoscopic array, the optics transmission, the angular spread of the Fresnel lens, the focal plane detector geometry, the quantum efficiency of the MAPMTs and the trigger electronics were simulated by a proper code. The image analysis and event reconstruction were performed with an "ad hoc" procedure on the Cherenkov images [3].

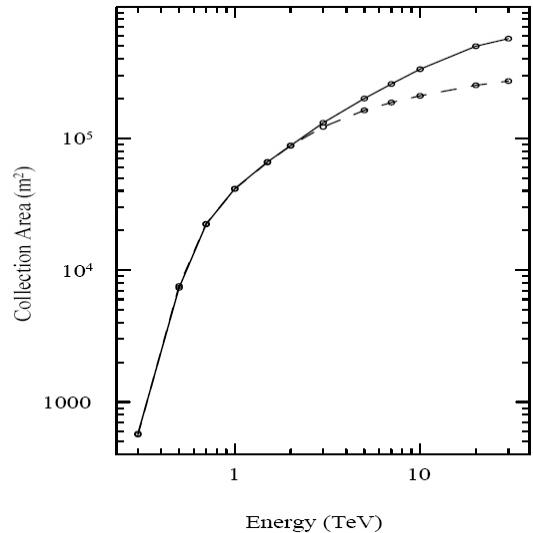

Figure 1: Collecting area of the GAW telescope array vs energy for on-axis Gamma Ray events. Solid line refers to a focal plane detector with a large FoV of 24°x24° while the dashed line refers to a small FoV of ~5°x5°.

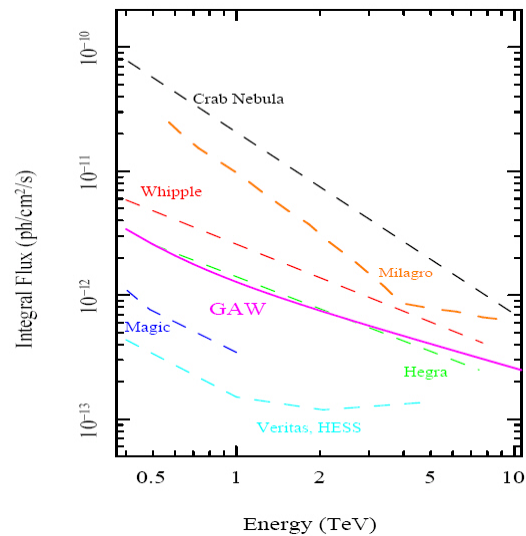

Figure 2: The sensitivity limit (5$\sigma$) detection of GAW (solid line) for Crab-like point sources in 50 hours observations. The flux of the Crab Nebula and the sensitivity of other TeV experiments (dashed lines) are also shown for comparison. The Milagro sensitivity is for 1 year of observation.

Fig. 1 shows the collection area vs. energy for mono-energetic gamma-ray events coming from a cosmic on-axis source at the zenith and with the constraints of trigger coincidence in all three telescopes. The solid line represents the collection area with a FoV of 24°x24°, while the dashed line shows, for comparison, the collection area with the reduced FoV of ~5°x5° . The advantage of a large FoV is even more striking at higher energies. The detection trigger rate for a Crab-like spectrum peaks at 0.7 TeV.

The main performance of GAW is summarized by its integrated flux sensitivity as function of the energy. Fig. 2 shows the sensitivity for a Crab-like point source in 50 hours observation with 5 sigma detection limit. For comparison, the flux of the Crab Nebula and the sensitivity of other TeV experiments are also shown.

## Summary

IACTs with large FoV will offer two important advantages: they will survey the sky for serendipitous TeV detections and, at the same time, will increase the IACT collection area, triggering events whose core is far away from the telescope axis and therefore improving the statistics of the high energy tail of the source spectra.

Presently, GAW is a R&D experiment made up of an array of three identical relatively small Cherenkov telescopes that will test the feasibility of a new generation IACT that joins large FoV and high flux sensitivity. Large FoV will be achieved by using refractive optics made of single side flat Fresnel lens of moderate size (2.13 m diameter). The focal camera will use the single photon counting mode instead of the charge integration mode widely used in the present IACT experiments. This detector working mode will allow us to operate with a very low photoelectron threshold and a consequent lowering of the energy threshold. The stereoscopic observational approach will improve the angular resolution, the cability of identifying gamma-ray induced showers and a good determination of the primary photon energy. In this way it will reach the necessary sensitivity to accomplish the foreseen scientific goals.

GAW is a collaboration effort of Research Institutes in Italy, Portugal and Spain. It will be erected in the Calar Alto Observatory (Sierra de Los Filabres - Andalucía, Spain). The first telescope will be settled within 2007.

## Acknowledgements

The GAW collaboration acknowledge the support by the CSIC, the grant by the Junta de Andalucía under project numbers FQM160 and P06-FQM-01392 (Spain) and the support of FCT - Fundacão para a Ciência e Tecnologia (Portugal).